# Applying Software Defect Estimations: Using a Risk Matrix for Tuning Test Effort

James Cusick
Wolters Kluwer
*j.cusick@computer.org*

## ABSTRACT

*Applying software defect estimation techniques and presenting this information in a compact and impactful decision table can clearly illustrate to collaborative groups how critical this position is in the overall development cycle. The Test Risk Matrix described here has proven to be a valuable addition to the management tools and approaches used in developing large scale software on several releases. Use of this matrix in development planning meetings can clarify the attendant risks and possible consequences of carrying out or bypassing specific test activities.*

## INTRODUCTION

Managing software system testing sometimes lands technical staff firmly between a rock and a hard place. On the one hand development overruns tend to cut into scheduled test time in order to meet product delivery commitments. On the other hand professional duty calls for extensive testing to uncover bugs before release. Applying software defect estimation techniques and presenting this information in a compact and impactful decision table can clearly illustrate to collaborative groups how critical this position is in the overall development cycle.

## DEVELOPMENT TUNING APPROACHES

In development shops with well defined processes sometimes the steps required to produce software can stretch out interminably. Not all software development efforts call for each step in a process to be carried out. In some cases a tailored development approach is crafted for each software project. DeGrace (1990) covers some of these approaches. One approach at NASA is called NASA-NMI-5330.1. This approach categorizes the planned software on 10 characteristics. The overall grade of the software can then be plugged into a Software Assurance Practices Grid indicating which quality practices are recommended.

Another development planning matrix described by DeGrace is DOD-STD-7925. In this approach a complexity assessment matrix is used to assign a complexity ranking to any software system. Based on that rating a minimum documentation grid is used to determine what documents should be produced with the software.

These planning tools provide some helpful guidance in development tuning for documentation and quality assurance activities but do not cover system testing activities. Further, these processes rely somewhat on subjective considerations such as if the design is complex or simple.

# SYSTEM TEST RISK MATRIX

During the management of some recent releases of software our team found a useful application of software development estimations in the selection of system test procedures. The Test Level Matrix proved helpful in clearly identifying which test activities would and would not be conducted. Specifically, this grid provided a simple communication tool for use with other development managers in negotiating the risks of not conducting certain tests or of cutting the test cycle. This risks were clearly enunciated in terms of defects not found by a shorter test cycle and thereby an increase in the number of defects delivered to the field. This application appears to build on the work noted above done by NASA and DOD but drives into the new area of system test tuning.

## *Process Integration of The Test Risk Matrix*

The Test Risk Matrix combines information from software development estimates and the software testing process to project the number of delivered defects for a software system release based on the intensity or level of testing carried out. Figure 1 below represents the information flow in the application of the Test Risk Matrix. Once system requirements are known and estimates of software size can be calculated the Test Risk Matrix can be applied. Placing the standard testing procedures in a table allows for a decision to be made on what level of testing will be conducted. Upon completion of testing improvements can be made to the understanding of the test process and future software estimations.

## *The Test Risk Matrix Details*

The Test Risk Matrix attempts to succinctly bring to light the risk, in terms of delivered defects, associated with choosing a particular testing schedule or coverage strategy. The contents of the Test Risk Matrix cover key characteristics of typical system test efforts. The first table is the Test Risk Matrix and the second table is the Test Scope Matrix. The Scope row shown in Table 1 is decomposed in Table 2 to provide test scalability and to maintain simplicity in Table 1. Each element of the Test Risk Matrix is defined below. Each element is either a "Standard" element or a "Custom" element. Standard elements generally do not change in value from application to application. Custom elements need to be calculated specifically for each application of the Test Risk Matrix.

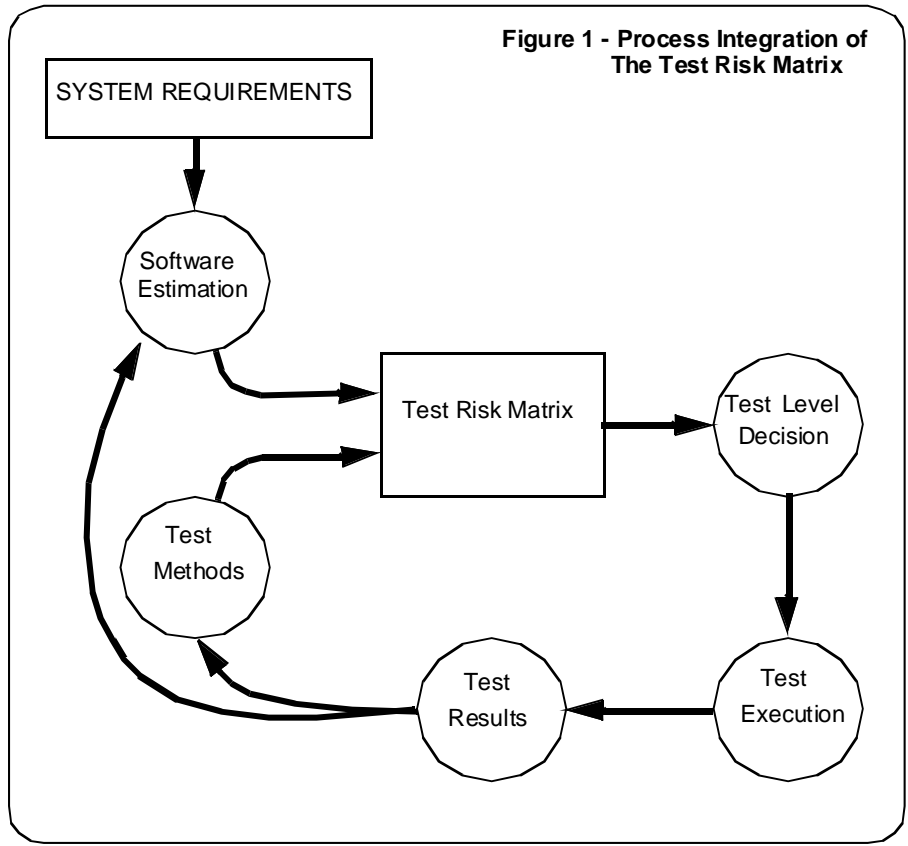

*Figure 1*

TEST SCOPE: Test process activities as expressed in terms of features, sanity suites, regression suites, destructive tests, stress tests, and field verification. Each scope incorporates different combinations of test process activities. See the Test Level Matrix below for specifics. STANDARD ELEMENT.
INTENSITY: An expression of how the test scope translates into actual test execution on a subjective scale. STANDARD ELEMENT.
ENVIRONMENT: Employ existing equipment or add new test infrastructure such as additional computing equipment, additional automated tools, and wider use of statistical analysis. STANDARD ELEMENT.
STAFF, STAFF WEEKS, CALENDAR WEEKS: The number of staff in technical head count, the number of combined staff weeks, and the number of calendar weeks needed for a given level of testing. CUSTOM ELEMENT.
PREDICTED DEFECTS: The number of defects projected to be in the software at the beginning of the system test phase. CUSTOM ELEMENT.
DEFECT REMOVAL EFFICIENCY (DRE): The percent of defects removed before delivery to the field. Normally this can never reach 100%. STANDARD ELEMENT.
DELIVERED DEFECTS (DD): The number of software defects delivered to the field after system test. Normally this can never reach 0. CUSTOM ELEMENT.

TEST RISK MATRIX

| TEST LEVEL | MINIMAL | LOW | MEDIUM | HIGH | EXTENSIVE |
|---|---|---|---|---|---|
| TEST SCOPE | A | B | C | D | E |
| INTENSITY | LIGHT | LIGHT | MEDIUM | STRONG | STRONG |
| ENVIRONMENT | Existing | Existing | Existing | Enhanced | Enhanced |
| STAFF | | | | | |
| STAFF WEEKS | | | | | |
| CALENDAR WEEKS | | | | | |
| PREDICTED DEFECTS | | | | | |
| DRE | 10% | 30% | 60% | 85% | 95% |
| DELIVERED DEFECTS | | | | | |

*Table 1 – Test Risk Matrix*

## *The Test Scope Matrix*

The Test Scope Matrix is used in conjunction with the Test Risk Matrix outlined above. The Test Scope element of the Test Risk Matrix relies on the construction and tuning of a Test Scope Matrix in order to gauge the test process steps considered for execution in the system test phase. This table takes as input standard testing activities as governed by development guidelines. These steps are arranged in the grid with inclusion and/or coverage characteristics. This gird can be expanded in either direction by adding new scope levels or "gray scales" along the top axis or by adding further test activities along the other axis. The Test Scope Matrix below in Table 2 represents a simplified version of an actual Test Scope Matrix. Potential expansion of the test steps could include usability testing, security testing, performance testing, storage testing, configuration testing, reliability testing, documentation testing. Definition of these and other test process steps can be found in Meyers (1979).

| SCOPE | A | B | C | D | E |
|---|---|---|---|---|---|
| Sanity | Yes | Yes | Yes | Yes | Yes |
| Features | Subset | Changed/New | Most | All | All |
| Regression | No | No | Minimal | Good | Complete |
| Stress | No | No | No | Good | Complete |
| Load | No | No | Minimal | Good | Complete |

*Table 2 – Test Scope Matrix*

# DERIVING ESTIMATES FOR THE MATRIX

In order to enliven this decision matrix requires estimations of the development effort of the system under consideration for test process tuning. Estimations can flow from a variety of sources and may come far in advance of the initiation of the system test phase or they can come "just-in-time" as system test begins. Estimates for new projects or for delta releases both find application in these decision tables.

## *Software and Staff Estimates*

Size metrics or functional metrics can be applied in the Test Risk Matrix. The key data elements required are the approximate number of defects expected to be found, the defect removal efficiency, and the amount of staff effort required to conduct the test phase. The worst case estimates should be arrived at first and then scaled down incrementally to fit the Test Risk Matrix. Staffing estimates often require heuristic methods of calculation but historical data on test case execution rates are the preferred method. Arriving at defect estimates can be done using historical data or if necessary use of simple mathematical formulas combined with industry average defect counts can serve the purpose just as well. Two such methods are briefly outlined in the following sections.

## *Defect Prediction Methods*

Historical data remains the superior method of defect prediction. Organizational vagaries such as staff skill and process sophistication can strongly influence the rate of defects in any software product. When historical defect data is not available a Function Point count or estimate can provide defect predictions for insertion into the Test Risk Matrix. Calculating Function Points for a retrofit requires taking the size of the system in LOC divided by a predefined source statement per function point value and adjusting for complexity (Jones, 1991).Taking the function point value and using published industry averages of defects per function point provides a very rough defect estimate for any software release. Other methods of defect prediction require only LOC and some defect adjustment parameters to calculate inherent and predicted defects (Musa, 1987). Once again if historical data is not available substitution of published averages can be used as a starting point for such values as defects per KLOC.

## *Defect Removal Efficiency or Test Effectiveness*

Defect Removal Efficiency (DRE), or its corollary Test Effectiveness in the test phase provide, the risk portion of the Test Risk Matrix. By displaying in the matrix the projected DRE of a set of test practices the full impact of phase tuning decisions can be visualized. Using the formula below provided by Card (1990) the DRE of past defect removal activities can be arrived at for use in the matrix.

$E = N/(N + S)$

where
$E$ = effectiveness of activity
$N$ = number of faults (defects) found by activity
$S$ = number of faults (defects) found by subsequent activities

# TEST PROCESS INPUT

In introducing the composition of the Test Scope Matrix earlier a variety of test procedures were listed. It is beyond the scope of this paper to discuss in detail these test types or how to arrive at which tests to conduct for a given software product. Based on the development standards of the organization, a catalogue of test types should be available for consideration when customizing the Test Scope Matrix for a software release. The Test Scope Matrix does not appear limited in the number of test types which it can include nor does it appear constrained in the levels of gray scaling which might be applied. This customization is best worked out by the test specialist in advance of using the Test Risk Matrix, however, it has been found that adjustments to the inclusion properties of some test activities within the Test Scope Matrix can be carried out swiftly during negotiations on selection of a testing level.

# AN EXAMPLE OF USING THE TEST RISK MATRIX

Putting the matrix to work on a simple example may assist in understanding the manner of its application. Consider a new software product estimated at 100,000 LOC or approximately 800 Function Points in its early stages. Estimates of defects could range from 650 to as much as 1400 at the time the system enters system test. Applying a sliding scale of DRE to these defect estimates, as the tables below indicate, brings us to the "bottom line" of software development and testing: Delivered Defects.

TEST RISK MATRIX

| TEST LEVEL | MINIMAL | LOW | MEDIUM | HIGH | EXTENSIVE |
|---|---|---|---|---|---|
| TEST SCOPE | A | B | C | D | E |
| INTENSITY | LIGHT | LIGHT | MEDIUM | STRONG | STRONG |
| ENVIRONMENT | Existing | Existing | Existing | Enhanced | Enhanced |
| STAFF | 2 | 2 | 4 | 5 | 5 |
| STAFF WEEKS | 6 | 12 | 32 | 60 | 80 |
| CALENDAR WEEKS | 3 | 6 | 8 | 12 | 16 |
| PREDICTED DEFECTS | 800 | 800 | 800 | 800 | 800 |
| DRE | 10% | 30% | 60% | 85% | 95% |
| DELIVERED DEFECTS | 720 | 560 | 320 | 120 | 40 |

*Table 3 – Test Risk Matrix Example*

TEST  SCOPE  MATRIX

| SCOPE | A | B | C | D | E |
|---|---|---|---|---|---|
| Sanity | Yes | Yes | Yes | Yes | Yes |
| Features | Subset | Changed/New | Most | All | All |
| Regression | No | No | Minimal | Good | Complete |
| Stress | No | No | No | Good | Complete |
| Load | No | No | Minimal | Good | Complete |

*Table 4 – Test Scope Matrix Example*

# FUTURE DIRECTIONS

The strength of the Test Level Matrix comes from the clear indicator of risk for each test level choice presented in the matrix. Work should be done on certifying the defect predictions and the success rate in delivering the projected number of defects. Further use of quantitative methods in aligning the test procedures could add rigor to the overall approach. An additional predictive value of software reliability might also find a place in the Test Risk Matrix along with Deliverd Defects. Finally, a link could be established from the development tuning grids mentioned above and the Test Risk Matrix. Such a link could clarify, for example, the additional test time required if certain development steps were clipped from the process, or the converse, how much test time can be saved by adding to the development stages.

# CONCLUSIONS

The Test Risk Matrix has proven to be a valuable addition to the management tools and approaches used in developing large scale software on several releases. Use of this matrix in development planning meetings can clarify the attendant risks and possible consequences of carrying out or bypassing specific test activities. With a minimum of preparation these tables can be constructed and modified for use on any project actively applying standard software metrics. In gearing up a software metrics program this matrix provides an early opportunity to demonstrate the applicability of software metrics in a decision support role.